\documentclass{article}
\usepackage{spconf,amsmath,graphicx}
\usepackage{subfigure}
\usepackage{algorithm}
\usepackage{algorithmicx}
\usepackage{algpseudocode}

\title{GraphTTS: Graph-to-sequence modelling in neural text-to-speech }
\name{Aolan Sun, Jianzong Wang\textsuperscript{*}\thanks{*Corresponding author: Jianzong Wang, jzwang@188.com}, Ning Cheng, Huayi Peng, Zhen Zeng, Jing Xiao}
\address{Ping An Technology (Shenzhen) Co., Ltd.}
\begin{document}
%
\maketitle
\begin{abstract}
This paper leverages the graph-to-sequence method in neural text-to-speech (GraphTTS), which maps the graph embedding of the input sequence to spectrograms. The graphical inputs consist of node and edge representations constructed from input texts. The encoding of these graphical inputs incorporates syntax information by a GNN encoder module. Besides, applying the encoder of GraphTTS as a graph auxiliary encoder (GAE) can analyse prosody information from the semantic structure of texts. This can remove the manual selection of reference audios process and makes prosody modelling an end-to-end procedure. Experimental analysis shows that GraphTTS outperforms the state-of-the-art sequence-to-sequence models by 0.24 in Mean Opinion Score (MOS). GAE can adjust the pause, ventilation and tones of synthesised audios automatically. This experimental conclusion may give some inspiration to researchers working on improving speech synthesis prosody.
\end{abstract}
\section{Introduction}
\label{sec:intro}

Text-to-speech (TTS), an integral part of the intelligent question answering system, has shown huge development over the past twenty years. Researchers try to achieve indistinguishable synthesised speeches within limited time and memory resources. Neural approaches, specifically sequence-to-sequence models, make the development of TTS to a higher level after the announcement of Google's Tacotron\cite{wang2017tacotron:} and Wavenet\cite{shen2018natural}\cite{oord2016wavenet} model. 

Although neural TTS has already shown competitive performance, prosody modelling is still a challenging task. A prosody embedding is first tried to be extracted from spectrograms and added as an additional input to the attention module\cite{wang2017uncovering}\cite{skerryryan2018towards}. Multi-head global style tokens are developed to represent audio's various speaking styles \cite{wang2018style}. These methods control the global style of synthesised audios, but the local speaking rhythm like pause, ventilation and tones is significant to the naturalness of synthesised audios. Younggun Lee \cite{lee2019robust} introduces temporal structure to enable fine-grained control of the speaking style of the synthesised speech. Ya-Jie Zhang \cite{zhang2019learning} introduces the Variational Autoencoder (VAE) to learn the latent representations of speaking styles which makes the sequence-to-sequence model easy for local style control. The above method solves the local prosody modelling of the speech synthesis process, but the technical barrier of the above methods is the requirement of manually selecting reference audio which may also introduce errors in the model. Some studies have attempted to extract rhythms from the textual content, but the addition of natural language processing (NLP) modules has greatly increased the complexity of the model.

Graph neural networks (GNNs) are connectionist models that capture the dependence of graphs via message passing between the nodes of graphs, which can do propagation guided by the graph structure of the input sequence \cite{scarselli2008graph}\cite{zhou2018graph}. The input text sequences can be processed as graph-structured to represent text contents by nodes embedding and represent the syntax and semantic connections as edge embedding. This can introduce text syntax information to the model if a graph encoder is leveraged in a sequence-to-sequence task. Some studies have already shown the advantages of applications of graph-to-sequence model on neural machine translation (NMT) task \cite{beck2018graph-to-sequence}\cite{bastings2017graph} where the graph embedding is analysed by dependency parsing methods by the SyntaxNet ToolKit published by Google group\cite{kong2017dragnn}\cite{song2018graph}. The speech synthesis task can be similarly modelled as a graph-to-sequence procedure where the graph embedding of text inputs needs to be designed and constructed because of the different mapping objective. So the contributions of this paper are: 
\begin{itemize}
\item A graphical text-to-speech model (GraphTTS) is proposed to map graph embedding to spectrograms, which leverages the GNN model in the field of neural speech synthesis for the first time; 
\item The Graph auxiliary encoder (GAE) module embeds prosody information into speech generation by analysing the input text, which makes prosody modelling an end-to-end procedure. 
\item A character-level graph embedding is constructed to map the input text to graph embedding from time-domain to space-domain with the semantic information embedded.
\end{itemize}

\begin{figure*}[t]
    \centering
    \includegraphics[scale=0.6]{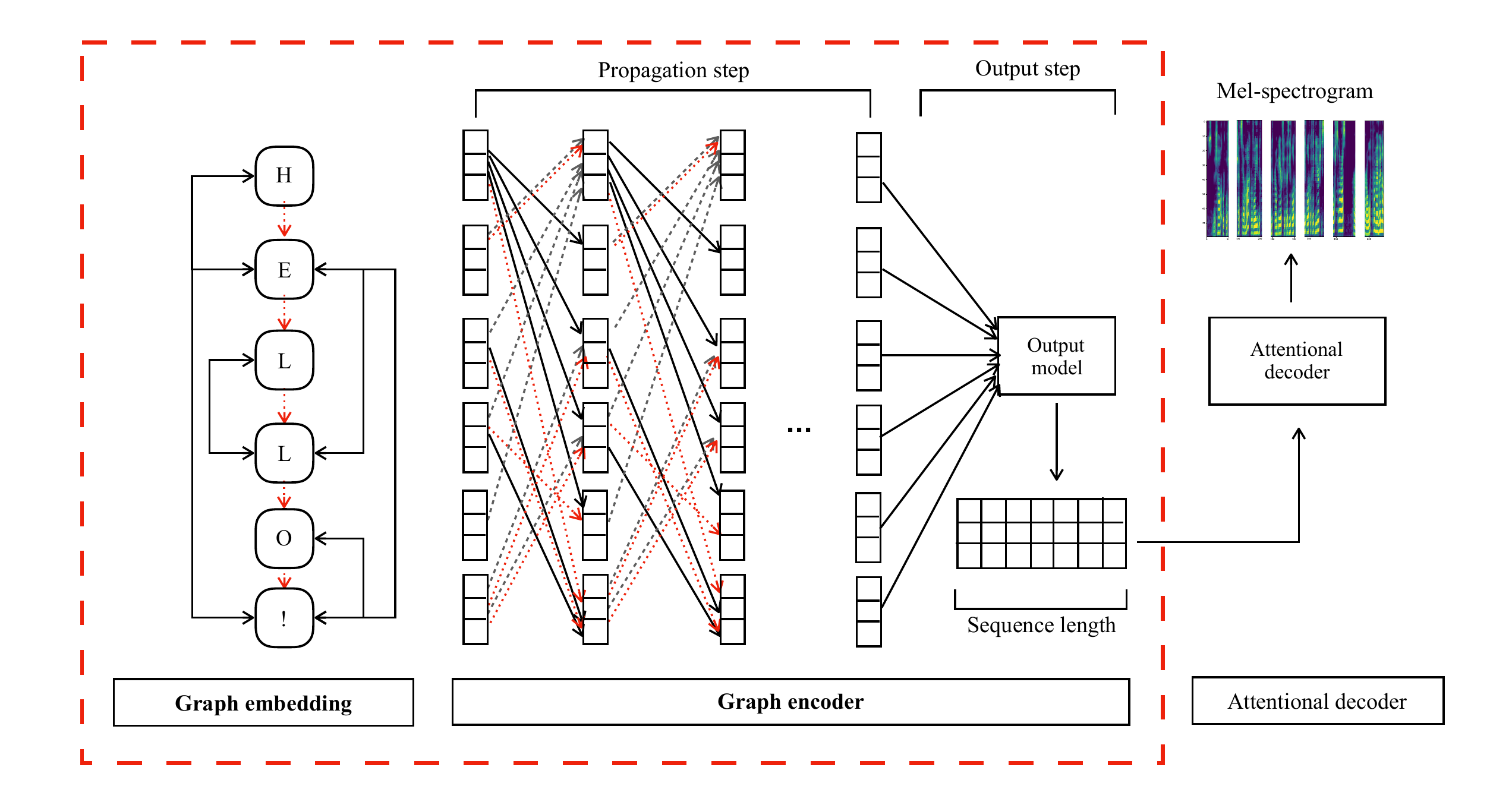}    \caption{GraphTTS model}
    \label{fig:G2STTS}
\end{figure*}

\section{Model architecture}
\label{sec:model}
In the sequence-to-sequence model \cite{sutskever2014sequence}, the recurrent neural network is normally used as the encoder module to map the input sequence to a variable-length output sequence. This process is mainly operated on the time domain to search for an alignment between text characters and corresponding frames of speeches. Graph neural network maps input sequence data into a space domain, which can represent syntax information as a graph embedding consumed by a graphical encoder to analyse prosody relationship among text characters. This section will describe the GraphTTS module in detail.

\subsection{GraphTTS}
\label{subsec:graphtts encoder}
The propagation procedure of GNN model is divided into two phases, the propagation step to compute node representation for each node and the output model to map node representations to encoder hidden states. A simple feed-forward neural network is used in the original graph neural network model. The variants utilise different aggregators to gather information from each node’s neighbours and specific updaters to update nodes’ hidden states in the propagation step. Gated graph neural network (GGNN) utilises a gated aggregator whereas graph convolutional network (GCN) uses a convolutional one. The difference between GCNs and GGNNs is that the learned filters depend on the Laplacian eigenbasis in GCNs whereas GGNNs diminish the Laplacian restrictions of GCNs and improve the long-term propagation of information flow across the graph structure. 

Since GCN models mainly deal with global graph inputs and make independent node-wise predictions, it can be used to extract prosody information from the graph embedding to incorporate syntax information into the model. \cite{bastings2017graph} shows the first application of graph convolutional network (GCN) on the neural machine translation task. \cite{marcheggiani2018exploiting} and \cite{vashishth2019incorporating} makes some improvements showing the feasibility of GCN model on the sequence-to-sequence task. 

GGNNs extends GCN models to sequential outputs by invoking gated units. GraphTTS utilises GCN or GGNN as the graph encoder which receives all nodes states as source states and delivers messages according to the adjacency matrix to aggregate messages to target nodes. The output of the propagation model is transformed by the output model to a graphical hidden state. This message passing process is shown in Figure \ref{fig:G2STTS}, where the edge embedding has three types, i) the directed edges (solid black lines), ii) the reverse edges (dashed grey lines), iii) the sequential edges (red dotted lines). This is somewhat like an alternative version of fully-connected neural networks that only the connecting edges participates in forward-backward propagation whereas edge weights of the others are represented by 0. \cite{beck2018graph-to-sequence} shows GGNN has better performance than the sequence-to-sequence models on the abstract mean representation (AMR) and neural machine translation (NMT) tasks, which shows the possibility of applying graph-to-sequence models in other sequence-to-sequence tasks. 

The graph embedding module shown in Figure \ref{fig:G2STTS} is a simple example where the nodes are represented by blocks and edges denoted by arrows. The remaining attentional decoder follows the same structure as Tacotron\cite{shen2018natural}, so this will not be described in detail here. 

\subsection{Graph auxiliary encoder}
\label{subsec: graphtts auxiliary encoder}
\begin{figure}[t]
    \centering
    \includegraphics[scale=0.6]{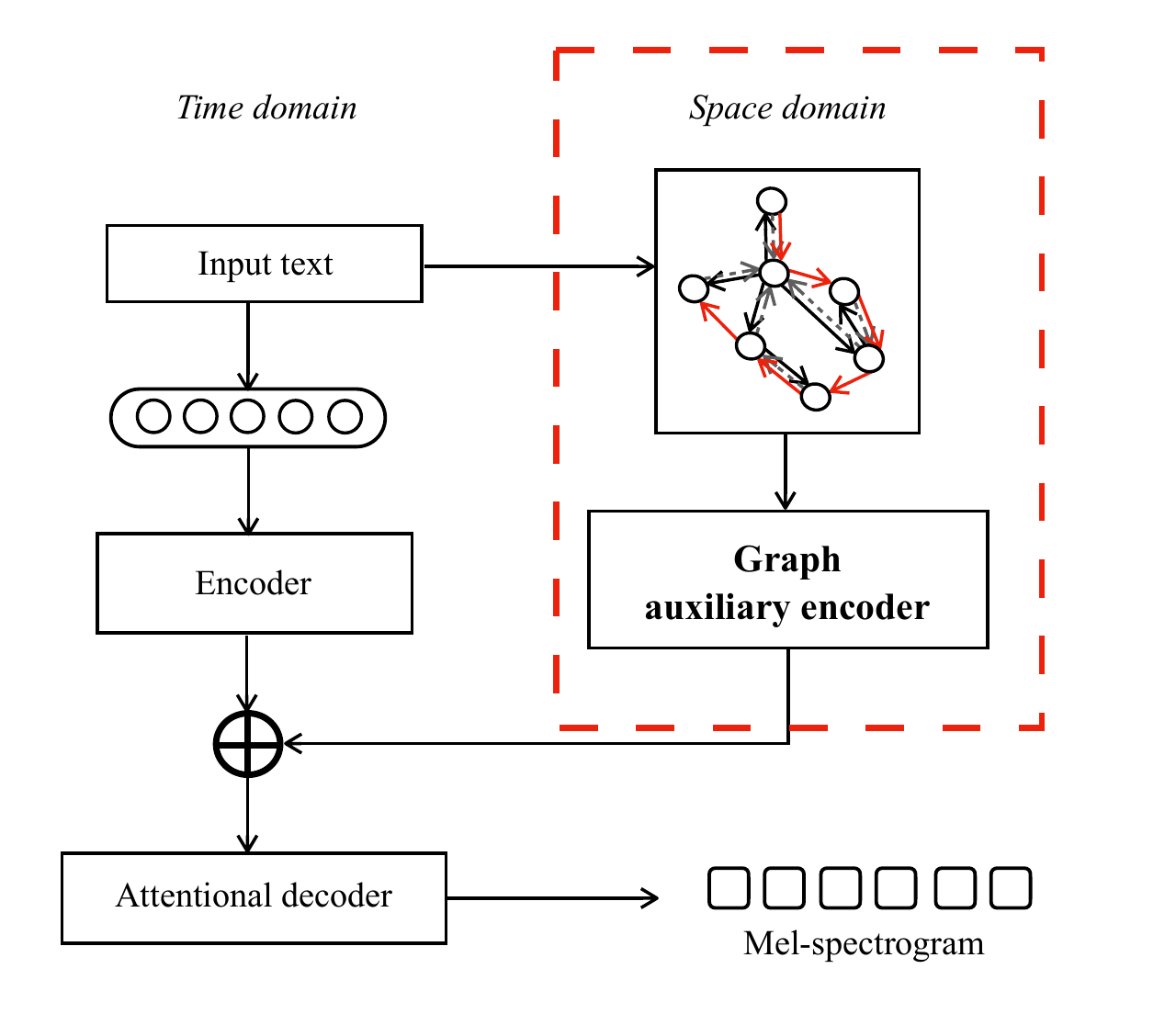}
    \caption{Graph auxiliary encoder}
    \label{fig:g2sref}
\end{figure}
Some researchers make some progress to use a reference encoder to capture prosody information from audios by several feature learning techniques \cite{wang2017uncovering}\cite{skerryryan2018towards}\cite{wang2018style}\cite{lee2019robust}\cite{zhang2019learning}. The above models can transfer the prosody from reference audio to the audios to be synthesised. This prosody information is mainly extracted from the audio prosodic patterns which are unrelated to the semantic contents of the input text. However, the encoder of GraphTTS can model the syntactic information, which is useful to the prosody adjustment. 

Therefore, Graph auxiliary encoder (GAE) in this section follows the same structure as the graph encoder of GraphTTS, to make the prosody modelling of speech generation an automated procedure. The main information flow of the model keeps the original structure of Tacotron, which guarantees the basic alignments of text characters and speeches. The transformation of input texts to graph embedding can be seen as a space-mapping procedure, projecting time-domain sequential inputs to a lower-level space-domain with syntactic information constructed. Since the main aligning module is reserved, the robustness of synthesised audios can remain whereas the prosody information is meanwhile extracted and embedded into the model.

The model architecture is depicted in Figure \ref{fig:g2sref}. The input sequence is first transformed into text embedding, which is then fed into the original encoder and outputs the encoder hidden states of the time domain. Meanwhile, the input text is transformed to graph embeddings, which is modelled by the Graph auxiliary encoder in the space domain. The GAE module is the same as the one in Figure \ref{fig:G2STTS} but with lower dimensions. The hidden state of the Graph auxiliary encoder is concatenated with the original encoder outputs as the queries to calculate the attention weights for model training. 

\section{Experiments}
\label{sec:experiment}

\subsection{Model configuration}
\begin{table}[t]
\small
\label{tab:GraphTTS hyperpara}
\vspace{0.5em}
\centering
 \begin{tabular}{p{0.45\columnwidth}p{0.3\columnwidth}}
    \hline
 Hyper-parameter & Tensor dimension  \\
    \hline
    Node embedding & 512-D \\
    Edge embedding & (E, 2, 3) \\
    GraphTTS-encoder & 512-D \\
    GAE & 128-D \\
    \hline
 \end{tabular}
\caption{Model configuration}
\end{table}
Experiments are conducted on a 24-hour female native speaker corpus LJSpeech and a 25-hour internal female Chinese dataset. Three types of graph encoders experiment. The parameter settings different from Tacotron are listed in Table 1, where $E$ in edge embedding denotes the number of edges of a graph, $2$ denotes two types of nodes, the source and target nodes, $3$ denotes the number of the types of edges.

\subsection{Data preparation}
\begin{figure}[t]
    \centering
    \includegraphics[scale=0.43]{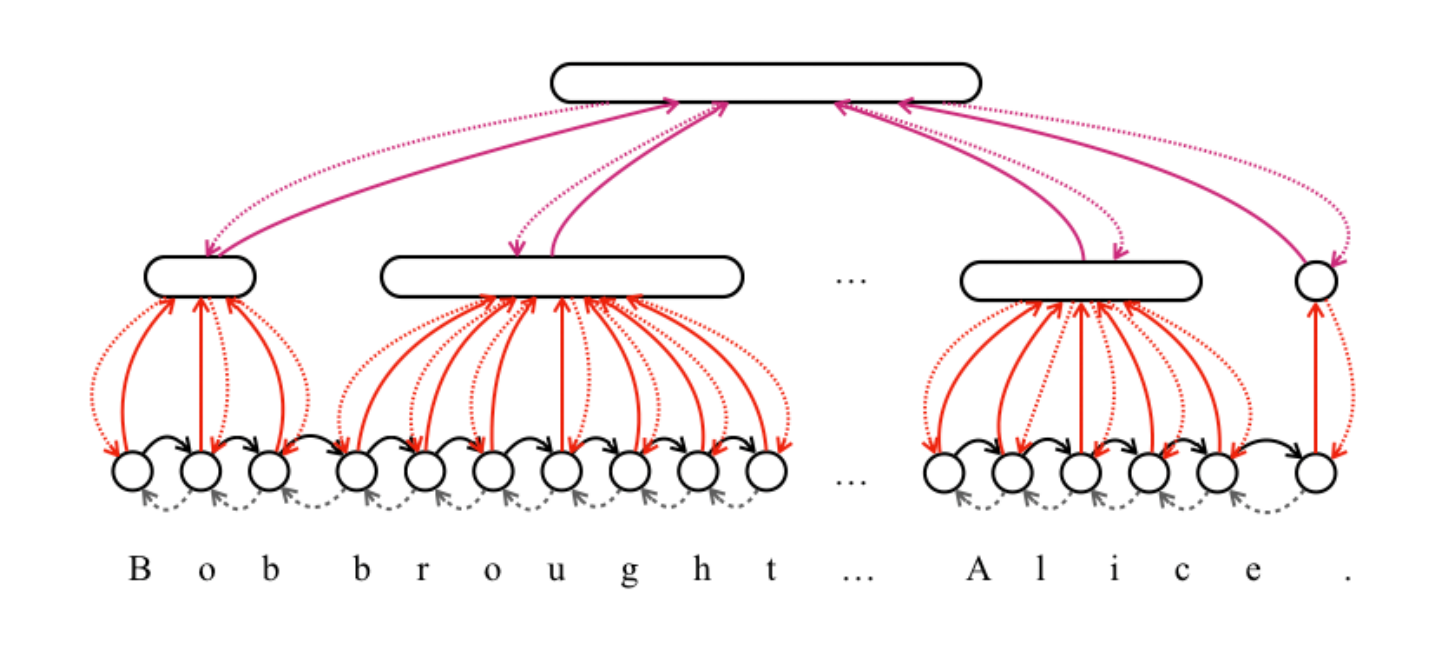}
    \caption{Character-level text-to-graph module}
    \label{fig:phoneme-based text-to-graph}
\end{figure}
The method used in \cite{beck2018graph-to-sequence} utilises dependency parsing trees to analyse word-level representations in the neural machine translation task. However, in natural speech synthesis, the phoneme of character is the basic unit to be modelled in neural speech synthesis, whose hidden states can be represented by the ones of raw characters when dealing with the end-to-end sequence-to-sequence model. The character-level information can be modelled as graph embedding shown in Figure \ref{fig:phoneme-based text-to-graph}. This structure forms the phoneme-level graph embedding of the input data, which is an alternative way to construct graph embedding

Characters of input texts are represented by nodes, and the adjacency connections among characters are modelled by edges. In Figure \ref{fig:phoneme-based text-to-graph}, the solid lines between neighbouring nodes represent the directed connections and the dashed lines represent the reverse one at the bottom character level. The connection between characters in a word is identified with the strong connection (represented by word-level red lines) and weak one among different words (represented by the sentence level purple lines). This structure forms the character-level graph embedding of the input data, which is an alternative way of dependency parsing tree methods with lower sparsity of nodes embeddings.

\subsection{Experiment I GraphTTS}
\begin{table}[t]
\small
\label{tab:GraphTTS model}
\vspace{0.5em}
\centering
 \begin{tabular}{ccc}
    \hline
    Dataset & English & Chinese \\
    \hline
    Tacotron1 & 4.17 & 4.0 \\
    Tacotron2 & 4.23 & 4.1\\
    Transformer TTS & 4.12 & 4.08 \\
    \hline
    GraphTTS GRU encoder & \textbf{4.47} & \textbf{4.31}\\
    GraphTTS LSTM encoder & 4.28 & 4.23\\
    GraphTTS GCN encoder & 4.17 & 4.12 \\
    \hline
    Human recording & 4.8 & 4.73 \\
    \hline
 \end{tabular}
\caption{GraphTTS model performance}
\end{table}
Three different types of graph neural network (GNN) models are experimented to evaluate in this section, gated recurrent unit (GRU), long-short term memory (LSTM) and graph convolutional network (GCN). The naturalness of synthesised audios were evaluated by Mean Opinion Score (MOS). 200 scorers scored anonymously through online surveys, of which 100 were native speakers and another 100 were bilingual. The MOS of the three state-of-the-art sequence-to-sequence models and human recordings are used as benchmarks to evaluate the naturalness performance of the proposed models. The experimental results are all shown in Table 2. 

It can be observed from Table 2 that the proposed three GraphTTS models outperform the three baseline models by a gap of approximately 0.2-0.3 MOS on both English and Chinese data set. Comparing the three graph encoder module, the MOS of the two GGNN encoders (LSTM and GRU) exceeds the GCN encoder and the GRU encoder gets the highest MOS. This result may be attributed to the long-term propagation property of GGNN models, which have a similar property of RNN modules whereas GCN is more analogous to CNN suitable for image or relation problems. 

\subsection{Experiment II Graph auxiliary encoder}
The experimental results of the graph auxiliary encoder model on the two data sets are shown in Table 3, where values of the metric represent the average value of English and Chinese data set. FR denotes the synthesis Failure Rate and PS denotes prosody score with a full score of 5. The metric PS is designed for measuring the richness, variability, and fullness of synthesised prosody. We use GraphTTS with GGNN module and GAE with GGNN in this experiment.

\begin{table}[t]
\small
\label{tab:exp21}
\vspace{0.3em}
\centering
 \begin{tabular}{p{0.4\columnwidth}p{0.1\columnwidth}p{0.1\columnwidth}p{0.1\columnwidth}}
    \hline
    Metric & MOS & FR & PS \\
    \hline
    Tacotron2 & 4.17 & 0.2\% & 3.8 \\
    \hline
    GraphTTS & 4.39 & 5\% & \textbf{4.5}\\
    GAE & \textbf{4.43} & \textbf{0.8}\% & 4.35\\
    \hline
    Human recording & 5.0 & $-$ & $-$ \\
    \hline
 \end{tabular}
\caption{Character-level model performance}
\end{table}

It can be seen that the graph auxiliary encoder (GAE) performs slightly better than GraphTTS but the synthesis failure rate of GAE is reduced to 0.8\% from 5\% of GraphTTS. This shows the robustness of the GAE model by keeping the time-domain modelling by the original sequence-to-sequence model. Besides, in the metric of prosody score, the GraphTTS score is higher, probably due to the effective propagation of feature representations in high-dimensional space, so that semantic information can be better embedded into synthesised speech.

\subsection{Experiment III Fine-tuning of hyper-parameters}
\begin{table}[t]
\small
\label{tab:TIME}
\vspace{0.5em}
\centering
 \begin{tabular}{ccc}
    \hline
    GraphTTS & iter=1 & iter=5 \\
    \hline
    MOS  & 4.47 & 4.53\\
    Model Convergence steps & 11k & 20k \\
    \hline
 \end{tabular}
\caption{Message passing iterations comparison}
\end{table}

An important hyper-parameter in message-passing process is $iter$, which represents the number of neighbouring nodes messages can pass. In particular, message only passes only the neighbouring nodes when $iter = 1$ whereas across five neighbouring nodes when $iter = 5$. Comparison experiments are conducted on it to tune the distance the message can pass through. The experiments are performed using GraphTTS on English data set for the best value of $iter$. The results in Table 4 show that the higher iterations, the higher MOS but slower convergence. But the improvement of $iter=5$ is not significant so that the computation of message passing across one neighbouring node can achieve reasonable results. 
 
\section{CONCLUSION}
\label{sec:conclusion}
In conclusion, the graph-to-sequence model is a very promising field in prosody modelling in neural text-to-speech. GraphTTS and GAE make a new exploration of the application of GNN model in the field of TTS. More GNN models such as GAT can be discussed and experimented to improve the prosody synthesised performance and naturalness of the synthesised audios. Besides, the word-level graph features may face bigger challenges because of its huge sparsity and high model complexity of the text-to-graph module. Deeper research can also be conducted on the incorporation of dimension-reduction methods or the effective word2vector approaches for appropriate word-level representations.

\section{ACKNOWLEDGEMENTS}
\label{sec:acknowledgement}
This paper is supported by National Key Research and Development Program of China under grant No. 2018YFB1003500, No. 2018YFB0204400 and No. 2017YFB1401202. Corresponding author is Jianzong Wang from Ping An Technology (Shenzhen) Co., Ltd.

\vfill\pagebreak

\bibliographystyle{IEEEbib}
\bibliography{refs}

\end{document}